# Time-Resolved In Situ Liquid-Phase Atomic Force Microscopy and Infrared Nanospectroscopy during the Formation of Metal−Organic Framework Thin Films

Laurens D. B. Mandemaker,[†] Matthias Filez,[†] Guusje Delen,[†] Huanshu Tan,[§] Xuehua Zhang,[§,‡] Detlef Lohse,[§,∥] and Bert M. Weckhuysen[*,†]

[†]Debye Institute for Nanomaterials Science, Utrecht University, Universiteitsweg 99, 3584 CG Utrecht, The Netherlands

[§]Physics of Fluids Group, Max Planck Center Twente, J. M. Burgers Centre for Fluid Dynamics, University of Twente, 7500 AE Enschede, The Netherlands

[‡]Department of Chemical and Materials Engineering, University of Alberta, Edmonton, Alberta T6G1H9, Canada

[∥]Max Planck Institute for Dynamics and Self-Organization, 37077 Goettingen, Germany

**S** Supporting Information

**ABSTRACT:** Metal−organic framework (MOF) thin films show unmatched promise as smart membranes and photocatalytic coatings. However, their nucleation and growth resulting from intricate molecular assembly processes are not well understood yet are crucial to control the thin film properties. Here, we directly observe the nucleation and growth behavior of HKUST-1 thin films by real-time in situ AFM at different temperatures in a Cu-BTC solution. In combination with ex situ infrared (nano)-spectroscopy, synthesis at 25 °C reveals initial nucleation of rapidly growing HKUST-1 islands surrounded by a continuously nucleating but slowly growing HKUST-1 carpet. Monitoring at 13 and 50 °C shows the strong impact of temperature on thin film formation, resulting in (partial) nucleation and growth inhibition. The nucleation and growth mechanisms as well as their kinetics provide insights to aid in future rational design of MOF thin films.

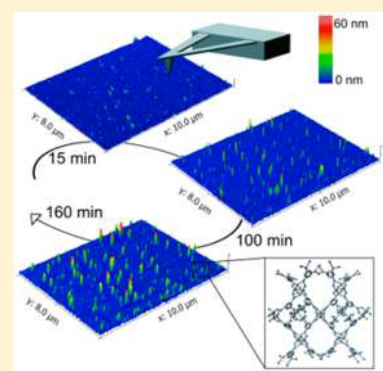

Metal−organic frameworks (MOFs) are versatile materials with high porosity, built up from metal clusters and organic linkers. The ability to vary linker, metal, and synthesis conditions leads to a great deal of flexibility to tailor the material properties and behavior, making them attractive for a diverse set of applications.[1,2] In particular, surface-mounted MOFs (SURMOFs) find use as smart membranes for gas sensing, separation, and storage, as well as photocatalytic coatings, photovoltaics, and electronics.[3−18] SURMOFs can be grown on Au using self-assembled monolayers (SAMs) as anchoring points for secondary building units from solution, serving as heterogeneous nucleation points. More particularly, film formation can be achieved via direct synthesis using a solution containing both metal and linker reagents or in a stepwise layer-by-layer (LbL) fashion in which the metal and linker solutions are separated. For both methods, the selected substrate and its functional groups play a crucial role during the film nucleation and growth as well as determining the final film properties.[20−22] Although the LbL method generally yields SURMOFs with low surface roughness and controlled growth coordination, the method consists of a multistep approach and is inconvenient to scale up compared to a "one-pot" direct approach. It is thus important to better understand the nucleation and growth mechanisms during direct SURMOF synthesis to more precisely control the thin film growth and properties. Atomic force microscopy (AFM) provides a powerful tool to monitor these materials[23] and has been used to study LbL nucleation and growth processes ex situ[22,24−26] and MOF-on-MOF crystal growth in situ over submicron length scales.[19,27,28] Yet, to the best of our knowledge, there are no reports of in situ AFM monitoring of the heteroepitaxial nucleation and growth of MOF thin films during direct synthesis.

Here, we report the nucleation and growth behavior of HKUST-1 thin films by real-time in situ AFM. More particularly, Cu-1,3,5-benzenetricarboxylic acid MOF (Cu-BTC) thin film formation is probed by liquid-phase AFM (Figure S1) at different synthesis temperatures. A 10 × 10 μm single spot on a 16-mercaptohexadecanoic acid (MHDA)-functionalized Au substrate in a metal-linker solution was continuously scanned (Figure 1a). The Au substrate was not varied for any experiments to ensure that the temperature was the only variable. Over time, the mixture of Cu precursor and BTC linkers nucleate and grow into HKUST-1 grains (Figure 1b).

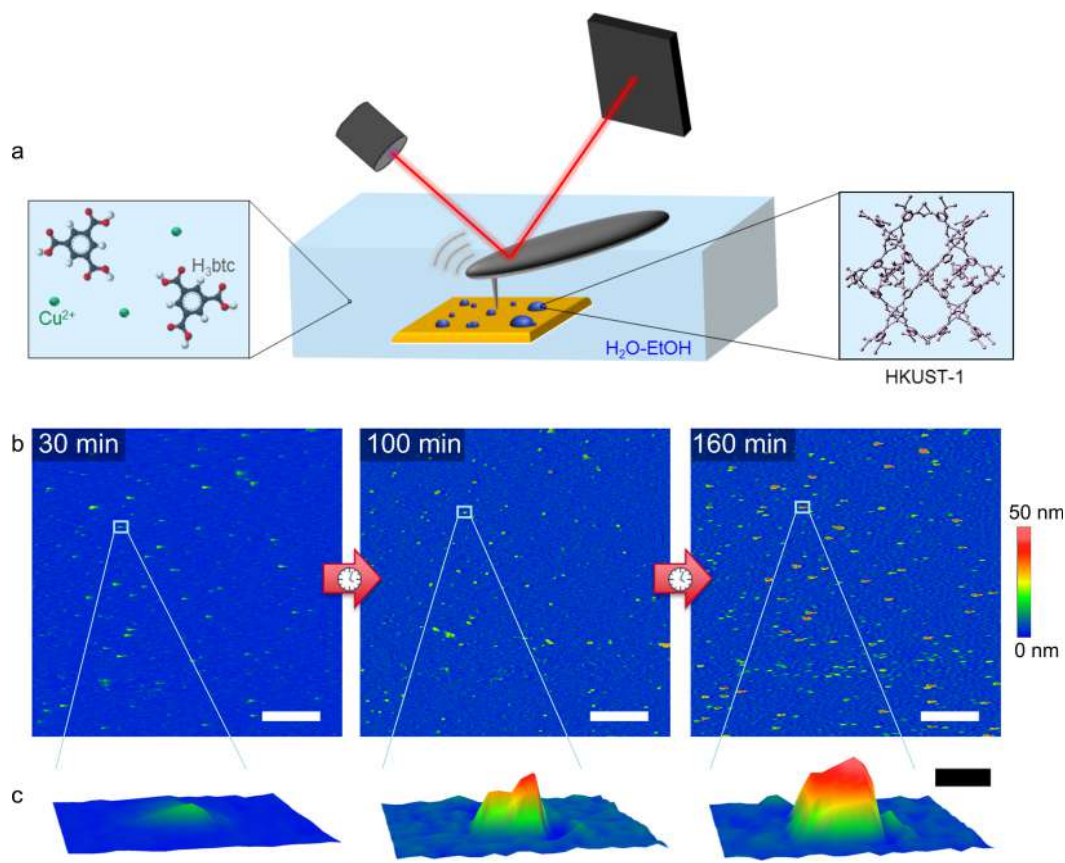

**Figure 1.** (a) Schematic of a MHDA-functionalized Au substrate continuously scanned with AFM over time (10 × 10 $\mu m^2$) while residing in a mixture of metal and organic linker solution; (b) AFM topographical scan of the same spot at 30, 100, and 160 min; and (c) zoom-in AFM topographical scan. This process was performed at 13, 25, and 50 °C (25 °C shown as an example), providing information on the influence of temperature when synthesizing a HKUST-1 thin film using a direct synthesis approach. The white scale bar is 1 $\mu$m, and the black scale bar 100 nm.

Scanning frequencies yielding approximately one time frame per 15 min were used, and time frames were plotted using the starting time of the AFM. To ensure the in situ AFM measurement itself did not interfere with the HKUST-1 growth, probes with a low force constant were used at low (~9 kHz) resonance frequencies. This was confirmed by scanning a HKUST-1 sample after 120 min of synthesis by comparing the settings above with more severe scanning settings (Figure S2). To study both the nucleation and growth behavior, we analyzed the in situ measurements using two different methods. First, we used a tailor-made script to label all features above a certain height threshold and track their height over time (see Supporting Information section 2).[29] Such a method has been often used to analyze the growth of surface nanobubbles and nanodroplets.[30] Rimer et al. reported comparable AFM studies on zeolites showing that the loss of resolution due to the change in tip geometry is less than 2.2% for the height (z-plane).[31] However, the x/y resolution tends to decrease. As a consequence of this result, only the height, and not the volume, of HKUST-1 grains was analyzed. Second, we used an open-source modular SPM program (Gwyddion) to filter all grains observed in the AFM image of one time frame and plot their heights as a "grain size distribution" over time.[32] In situ AFM maps and the selected grains can be found in Figure S3 for experiments at temperatures of 13, 25, and 50 °C. The height of all grains was plotted over time (Figure S4). To test if similar behavior was observed in other regions of the sample, two more spots were scanned postsynthesis (17 h) in Figure S5. As can be observed in Figure 1b, the heterogeneous nucleation and subsequent growth are of the Volmer−Weber type, reported previously in homoepitaxial growth studies and similar to what is seen in the in situ experiments.[19,22,25,28] Over the time scale of the experiment, nucleated HKUST-1 seeds grow into distinct 3-D islands on the SAM/Au substrate, rather than forming a uniform 2-D film. Such island formation is suggested to affect the eventual film roughness and uniformity and therefore ultimately relies on the observed nucleation and growth phenomena. The growth rates of individual grains could be derived from a linear fit of their height versus time. Boxplots of all growth rates at each temperature are shown in Figure 2a.

The median rates show faster growth rates at higher temperatures, which is the expected behavior for this material. The growth rates at 13 °C are low, with a median of 6.6 nm· $h^{-1}$. Also, some grains do not grow at all. The median growth rate almost doubles at 25 °C (11.4 nm·$h^{-1}$), and increasing the temperature even further to 50 °C results in a median growth rate of 15.6 nm·$h^{-1}$. A 3-D representation of the grain height distribution over time (for 25 °C) can be found in Figure 2b. Similar plots for 13 and 50 °C are found in Figure S6. Note that these are distributions of the entire 10 × 10 $\mu m^2$ AFM frame and not exclusively the selected grains from Figure S2. Nucleation plots are derived by plotting the number of grains over time (Figure 2c). Similar to the observed growth behavior, the synthesis temperature of 13 °C did not display an increase in the number of grains during the monitored time frames. The synthesis at 25 °C shows an increasing amount of nuclei over time, even after these 2.5 h of synthesis time. Because this keeps occurring, not only the higher, well-defined grains grow rapidly but also a background layer of HKUST-1 material— surrounding the larger grains—is formed, which we define as a HKUST-1 "carpet". Clearly, as shown in Figure 2c, the nucleation behavior at 50 °C is completely different. After only two time frames does the nucleation of new particles stop. Examining these maps ($t$ = 20 and 36 min, Figure 2d) reveals the formation of circular-like patches on the surface, and the formation of these patches is on the same time scale at which nucleation stops. These patches are very thin, having an approximate height of 2 nm (Figure 2e), which corresponds to the reported height of a MHDA layer.[33] This is explained by partial desorption of the MHDA from the Au substrate, leading to remaining islands of SAMs, instead of a monolayer. Although the Au−S bond is known to withstand such temperatures, it has been shown before on gold nanostars in aqueous solution that a partial desorption occurs around such temperatures.[34] To confirm the patches to be MHDA, AFM was used to study a Au, a fresh MHDA/Au, and a MHDA/Au substrate aged in an ethanol−H$_2$O mixture at 50 °C for 17 h, on multiple spots (Figure S7). Only the last substrate showed similar formation of patches, confirming the partial desorption of MHDA into solution at these elevated temperatures.

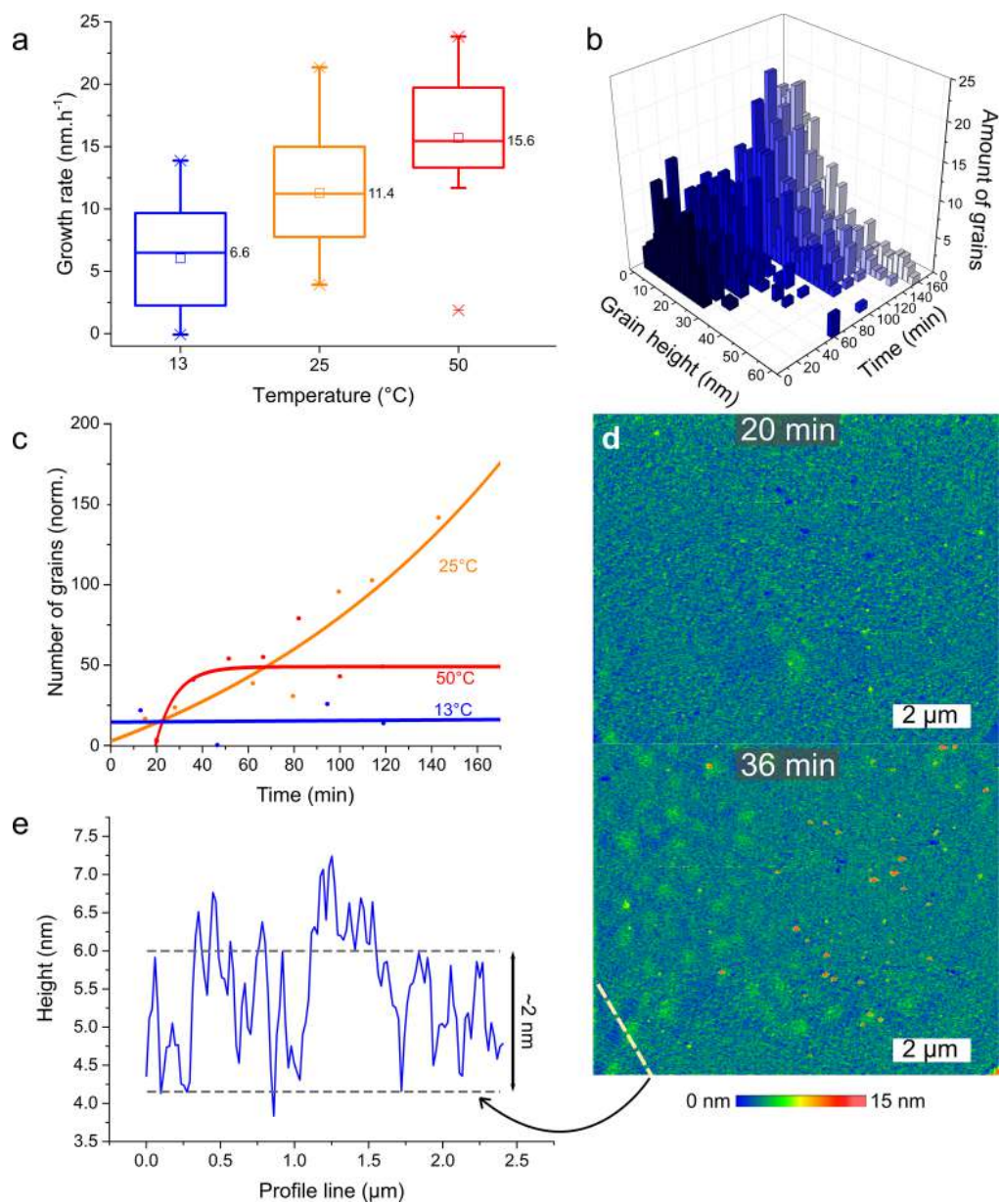

**Figure 2.** (a) Growth rates of all grains represented in boxplots at each temperature. (b) Grain size distribution (heights vs counts) as a function of time for HKUST-1 nucleation at 25 °C; plots for 13 and 50 °C can be found in Figure S6. (c) Total amount of grains plotted over time for different temperatures (13 °C, blue; 25 °C, orange; 50 °C, red); for 50 °C, the nucleation is quenched at $t$ = 36 min. (d) AFM maps at 50 °C show circular patches, which are formed between 20 and 36 min. (e) Height profile for the dashed line over two patches in (d); the difference in height is approximately 2 nm, reported before as the thickness of a MHDA layer.

To interrogate the chemical nature of the deposited HKUST-1, a sample synthesized at 25 °C was measured with XRD and SEM-EDX (Figures S8 and S9). Typical HKUST-1 peaks are present in the diffractogram, where the presence of the (111), (200), and (220) peaks highlights the nonuniform orientation of the film.[35] SEM-EDX maps and corresponding energy dispersive curves shows carbon, oxygen, and copper to be present on larger grains, corroborating the presence of HKUST-1, although SEM-EDX is not sufficiently sensitive to detect Cu or C in the carpet or small grains.

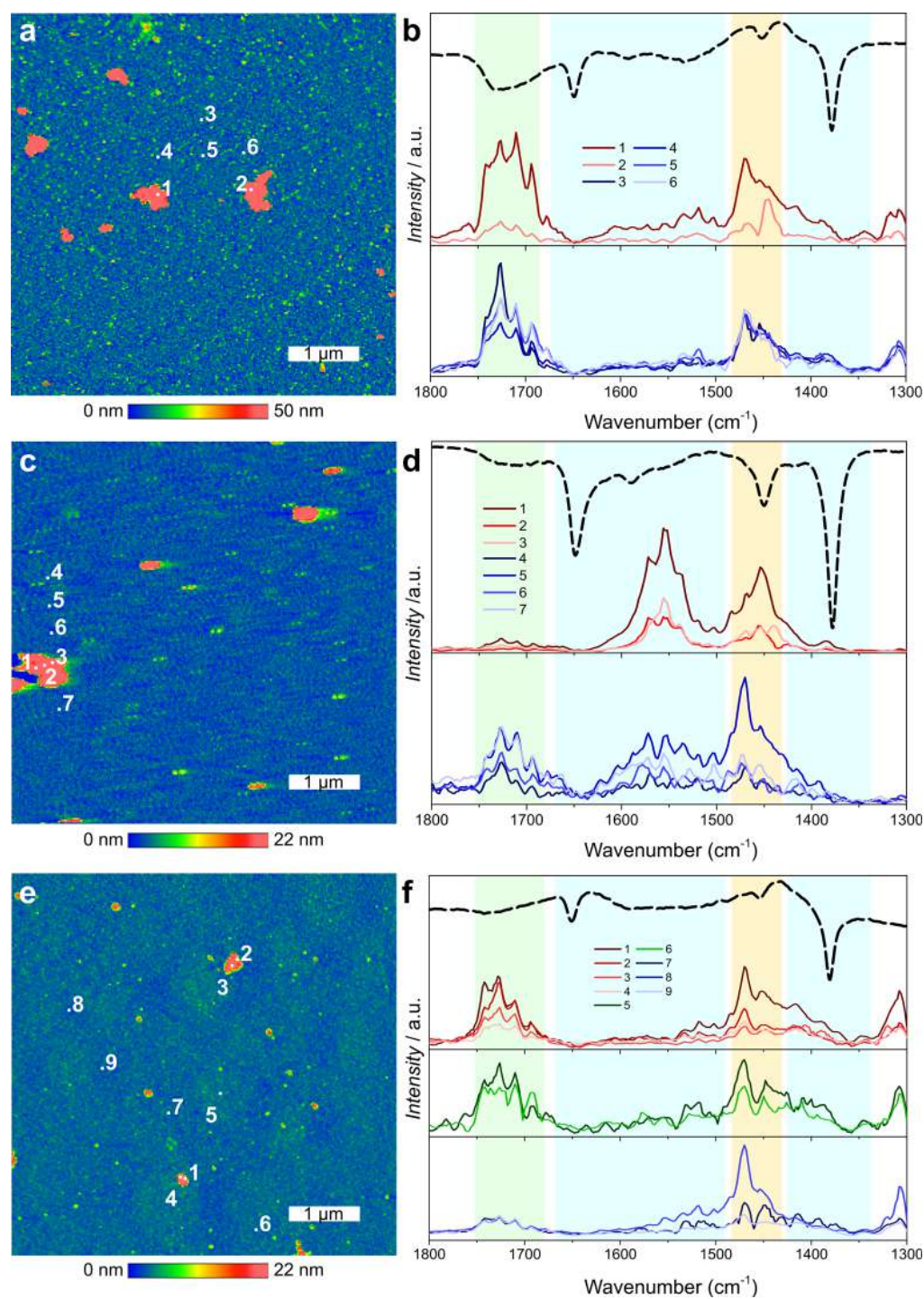

**Figure 3.** IR nanospectroscopy with AFM topography maps (a,c,e) and corresponding PiFM point spectra (b,d,f) for samples grown at 13, 25, and 50 °C for 17 h, respectively. The spectra are grouped related to their position: on grains (red), on carpet (blue), and on the formed patches (green). IRRAS spectra (dashed) are included for each temperature, share the same intensity scale, and offer a bulk comparison to the point spectra.

IR spectroscopy provides chemical information on the coordination of BTC linkers to the metal nodes of the deposited material. Bulk infrared reflection−absorption spectroscopy (IRRAS) is ideal for the characterization of thin film materials grown on reflective (gold) substrates. IRRAS spectra of samples synthesized at different temperatures (17 h) are shown in Figure 3 (dashed lines).

These spectra show characteristic sharp bands at ∼1380 and 1650 cm$^{-1}$, which represent the symmetrical and asymmetrical COO$^-$ stretches of the BTC linker coordinated to the copper cluster, respectively. The turquoise boxes in Figure 3 cover the region including the asymmetrical COO$^-$ stretches around 1650 and 1550 cm$^{-1}$ (vide infra) as well as the symmetric COO$^-$ stretch around 1380 cm$^{-1}$, which we refer to as the "Cu-BTC" bands, as they are generally accepted by literature.[36−38]

The band at ~1450 cm$^{-1}$ is assigned to the BTC benzene breathing vibration (yellow box). The broad (and less intense) band at ~1700−1750 cm$^{-1}$ belongs to the −COOH stretching vibration of uncoordinated MHDA (green box), as corroborated by the IRRAS reference spectrum of bare MHDA/Au in Figure S10. To gain complementary chemical information with nanometer resolution to link IRRAS and in situ AFM data, nanoinfrared spectroscopy, in the form of photoinduced force microscopy (PiFM), is used. The PiFM technique is able to avoid the diffraction limit by measuring the physical interaction of a laser-induced dipole and its mirror image in the Au-coated AFM tip, resulting in an IR spectra with a spatial resolution down to the nanometer level.[38−40] In addition to bulk IRRAS spectra, Figure 3 displays AFM topography maps and nano-IR point spectra for SURMOFs grown at 13 °C (Figure 3a,b), 25 °C (Figure 3c,d), and 50 °C (Figure 3e,f). Spectra are plotted together based on location (red, on a grain; blue, on the carpet). Because of a decrease in laser power between 1660 and 1620 cm$^{-1}$, most of the asymmetrical COO$^-$ stretch was not observed. Also, as detailed in the SI (section 4), additional effects can perturb the peak features recorded by PiFM, as compared to IRRAS. Nevertheless, the combination of IRRAS (mm-scale) and PiFM (nm-scale) respectively provides averaged, bulk chemical information complemented by local spectral snapshots of the formed islands and carpet at given synthesis conditions.[40]

The sample synthesized at 13 °C (Figure 3a,b) shows weak Cu-BTC bands and relatively strong MHDA bands for IRRAS as well as for PiFM measurements on both the grains and the carpet. This demonstrates the minor formation of HKUST-1 clusters, even after 17 h, as we already observed during our in situ AFM measurements (Figure 2a,c). The synthesis at 25 °C yields more HKUST-1 material, as is reflected in the Cu-BTC band intensities (blue region) in Figure 3d. The dominant band at ~1550 cm$^{-1}$ partially originates from the HKUST-1 asymmetrical COO$^-$ stretch and has been reported before on Cu(NO$_3$)$_2$-based HKUST-1.[41] The MHDA band is less intense in the spectra (IRRAS and PiFM) measured on the carpet and is hardly observed in point spectra taken on grains. This shows the large amount of HKUST-1 formed on these spots as the probing depth of both techniques was insufficient to measure the MHDA signal through the SURMOF. The carpet still has a well-defined MHDA band, but it is less intense than the band observed in the 13 °C synthesis. Again, this confirms our AFM measurement, where we observed continuous nucleation at 25 °C (Figure 2c). Finally, the synthesis at 50 °C shows lower intensity for both Cu-BTC and MHDA bands (Figure 3e,f). In this case, we split the carpet into "patch" (green spectra) and "off-patch" (blue spectra) regions. Both grains and patches show similar MHDA and Cu-BTC band intensities, with relatively less MHDA than for 13 °C but more than for 25 °C. Here we observe a minor MHDA band intensity on the off-patch points, verifying that MHDA desorbed at these spots and substantiating our claim that the formed patches are due to SAM desorption (Figure 2). The combined information shows that the synthesis at 13 °C yields very little HKUST-1 due to slow nucleation and growth, while at 25 °C, more HKUST-1 is formed as both islands and (disordered) carpet, and finally, at 50 °C, little HKUST-1 is formed due to detachment of the SAM.

In summary, real-time in situ liquid-phase AFM has been performed on the formation of SURMOFs at different temperatures, using HKUST-1 as a showcase. Besides the growth of larger HKUST-1 islands, which were individually studied, we found that the direct synthesis method also yielded a thin layer—termed carpet—surrounding the more rapidly growing grains. Combining this in situ approach with a postsynthesis PiFM analysis, we show that both grains and carpet are HKUST-1, and their growth is strongly influenced by the synthesis temperature. The powerful combination of these techniques with high spatial resolution offers promising perspectives for studies on growth mechanisms of similar SURMOFs or even other thin film materials.

## ■ EXPERIMENTAL METHODS

AFM measurements were performed on an NT-MDT NTEGRA Spectra system using NSG01 probes (ex situ, in air, $F$ = 5.1 N/m) in tapping mode with a resonance frequency of 150 kHz or Bruker SNL-D tips (in situ, in liquid, $F$ = 0.06 N/m) in tapping mode with a resonance frequency of approximately 9.5 kHz. PiFM AFM-IR measurements were performed on a VistaScope instrument at Molecular Vista in San Jose, CA. Topography and IR measurements were performed in tapping mode using PPP-NCHAu tips ($F$ = 42 N/m, resonance frequency = 330 Hz), applying a Bloch QCL laser ranging from 1300 to 1800 cm$^{-1}$ (with a noticeable dip in laser power in the range of 1620−1660 cm$^{-1}$). Additional liquid-AFM, IRRAS, XRD, SEM-EDX, and script details can be found in the SI.

## ■ ASSOCIATED CONTENT

### Ⓢ Supporting Information

The Supporting Information is available free of charge on the ACS Publications website at DOI: 10.1021/acs.jpclett.8b00203.

Additional experimental, spectroscopic, and analytical details, as well as additional data at different synthesis temperatures (PDF)

## ■ AUTHOR INFORMATION

### Corresponding Author
*E-mail: b.m.weckhuysen@uu.nl.
### ORCID 
Detlef Lohse: 0000-0003-4138-2255
Bert M. Weckhuysen: 0000-0001-5245-1426
### Notes
The authors declare no competing financial interest.

## ■ ACKNOWLEDGMENTS

The authors kindly acknowledge Molecular Vista for allowing us measurement time on the VistaScope instrument and specifically Katie Park for her help during measurements and Jochem Wijten (Utrecht University) for performing the SEM-EDX measurements. This work was supported by The Netherlands Center for Multiscale Catalytic Energy Conversion (MCEC), an NWO Gravitation program funded by the Ministry of Education, Culture and Science of the government of The Netherlands, by the European Research Council (ERC) Advanced Grant (No. 321140), and by a European Union's Horizon 2020 research and innovation program under the Marie Skłodowska-Curie grant agreement (No. 748563).

## ■ REFERENCES